\documentstyle[11pt]{article}

\newcommand{\be}{\begin{equation}}
\newcommand{\ee}{\end{equation}}
\newcommand{\ba}{\begin{array}}
\newcommand{\ea}{\end{array}}
\newcommand{\bc}{\begin{center}}
\newcommand{\ec}{\end{center}}
\newcommand{\bi}{\begin{itemize}}
\newcommand{\ei}{\end{itemize}}

\newcommand{\disregard}[1]{{}}

\def\bild#1\over#2{\mathrel{\mathop{\kern0pt #1}\limits_{#2}}}

\begin{document}

\centerline {{\bf IDEAL ANYONS\footnote{Mesoscopic
Quantum Physics -
Les Houches, Session LXI, 1994 -
E. Akkermans, G. Montambaux and J.-L. Pichard, eds. -
North- Holland, Amsterdam, to be published} \rm}}
\vskip 1cm

{\centerline {\bf St\a'ephane OUVRY \rm
\footnote{\it  and
LPTPE, Tour 16, Universit\'e Paris  6 / electronic e-mail: OUVRY@FRCPN11}}}
\vskip 1cm
{\centerline {Division de Physique Th\'eorique \footnote{\it Unit\a'e de
Recherche  des
Universit\a'es Paris 11 et Paris 6 associ\a'ee au CNRS},  IPN,
  Orsay Fr-91406}}

\vskip 1cm
{ Abstract : A general introduction to  the anyon model
(braid group, Chern-Simons Lagrangian and Aharonov-Bohm Hamiltonian
formulations) is given. A review follows on exact results and possible
ways of getting  additional information, as mean field
approach, perturbation theory, and projection on the lowest Landau level
of an external magnetic field. }

\vskip 3cm


IPNO/TH 94-76 (November 1994)

\vfill\eject

In 2+1 dimensions,  intermediate
statistics, which interpolate continously between Bose and Fermi
statistics, do exist [1]. Usually, quantum wavefunctions
$\psi(\vec r_1,\vec r_2)$ of two
undistinguishable particles are such that $\psi(\vec r_1,\vec r_2)=\pm
\psi(\vec r_2,\vec r_1)$, for
bosons or fermions. For anyons,
$\psi'(\vec r_1,\vec r_2)=e^{\pm i\pi\alpha}\psi'(\vec r_2,\vec r_1)$ is the
new rule. If $\alpha$ is an even integer, one has
Bose statistics, and if $\alpha$ is an odd integer, one has Fermi
statistics. Otherwise, the statistics is intermediate. How is this
possible ? Consider [2] the configuration space $\{\vec r_1,\vec
r_2,...,\vec r_N\}$ of $N$ identical particles
$E_N={R^{2N}-D^N\over S_N}$, where $D_N$, the diagonal
$\{\vec r_1,\vec
r_2,...,\vec r_N\|\vec r_i=\vec r_j\}$,  has been substracted,
and, since the particles are indistinguishable, where all the configurations
which differ by a permutation in $S_N$ have
been identified. One wants to characterize, in the configuration space,
 the interchange properties of
the particles, and, more generally, the topology of the space of loops of
particles around each other. This information is contained in the first
homotopy group of $E_N$, $\Pi_1(E_N)$, where
two loops are homotopic if and only if
one can continuously deform one into the other. In dimension 2,
$\Pi_1(E_N)=B_N$ is the braid group for $N$ particles, whereas
in dimension 3 and above, it is trivially $S_N$.
It is only in dimension 2
that the notion of winding around each other for 2 particles has a unambiguous
meaning.
 In
other
words, the plane is punctured at the location of the particles and loops of
particles around each other may be not contractible to a point, due to
 the topological obstructions materialized by these punctures.
The generators of the braid group are the transpositions
$T_i$,
which interchange particles $i$ and $i+1$. They satisfy to the algebra
$T_iT_j=T_jT_i$ if $|i-j|>1$, $T_iT_{i+1}T_i=T_{i+1}T_iT_{i+1}$, and,
contrary to the usual $S_N$ relation, one has
$T_i\ne T_i^{-1}$, which
implies that exchanging 1 and 2 clockwise or anticlockwise is not
innocent. Since these two exchanges are parity images, parity is broken,
and
the configuration space is multivalued : a given
configuration, in this example $\{\vec r_2,\vec r_1\}$, can be attained
in several inequivalent ways, from a given initial configuration  $\{\vec
r_1,\vec r_2\}$. The parameter $\alpha$ is defined as the phase factor
$\exp(\pm i\pi\alpha)$ associated to each generator $T_i$ (or $T_i^{-1}$)
in the case of one dimensional unitary
representations of $B_N$.

How can one realize this
non trivial phase factor in usual quantum mechanics? Bosons and fermions
are described by a free Hamiltonian with symmetric or
antisymmetric
boundary conditions on the wavefunctions.
What is the proper generalization of these boundary conditions to anyons?
Consider the free Hamiltonian $H'_{N}=\sum_i
\vec {p}_i^2/2m$ for
$N$ free
particles in the plane,
and impose to their wavefunction $\psi'(\vec r_1,\vec r_2, ... ,
\vec r_N)$ to be multivalued in the
configuration space according to $\exp(i\alpha\sum_{k<l}\theta_{kl})$,
where $\theta_{kl}$ is the relative
angle between particules $k$ and $l$.
In the case of a simple clockwise or anticlockwise transposition
$\theta_{kl}\to
\theta_{kl}\pm\pi$, one recovers the desired phase factor
$\exp(\pm i\pi\alpha)$. In quantum mechanics, one usually prefers to
work with
monovalued wavefunctions. There is an equivalent way to describe the
model, simply by trading off the multivaluedness of the free
$N$-body wavefunctions against
monovalued (bosonic or fermionic) but interacting $N$-body wavefunctions.
Simply gauge
transform
\be\label{gt} \psi'(\vec r_1,....,\vec r_N)=
e^{i\alpha\sum_{k<l}{\theta_{kl}}}\psi(\vec r_1,....,\vec r_N)\ee
the free Hamiltonian $H'_N$ to get
\be\label{H} H_N={1\over 2m}\sum_{i=1}^N(\vec p_i-\alpha\vec A(\vec r_i))^2\ee
with $\vec A(\vec
r_i)=\vec{\partial}_i\sum_{k<l}\theta_{kl}=\sum_{j\ne i}{\vec k\times\vec
r_{ij}\over r_{ij}^2}$
($\vec k$ the unit vector perpendicular to the plane, $\vec r_{ij}=\vec
r_i-\vec r_j$). The potential vector $\vec {A}(\vec r_i)$ has a simple
physical interpretation as a sum of Aharonov- Bohm   ($A-B$)
interactions [3] which couple a particule of charge $e$  at position $\vec r_i$
to infinitely thin vortices of flux
$\phi$ at position $\vec r_j$, with
the dimensionless  $A-B$  coupling
$\alpha=e\phi/2\pi= \phi/\phi_o$ -$\phi_o$ is the flux quantum.
The system is periodic in $\alpha$, with period 2, because the
shift $\alpha\to \alpha+2$ can be undone by a regular gauge
transformation (\ref{gt}) with parameter $\alpha=2$, which does not
affect the symmetry of the wavefunctions (symmetric or
antisymmetric). In the sequel, one will take $\alpha\in [-1,1]$,
with by convention $\alpha=0$ Bose statistics -i.e. one works with
bosonic wavefunctions-, and therefore $\alpha=\pm 1$ Fermi
statistics. The interpolations Bose-Fermi when
$\alpha : 0\to 1 $ or $0\to-1$, are equivalent since the sign of the flux tube
has no
physical meaning. However, when the anyons are coupled to an external
magnetic field, it defines
an orientation to the plane, and the sign of $\alpha$ becomes
a physical observable.
It is not a surprise that, statistics being a
purely quantum concept,  pure quantum  $A-B$ interactions,
without any classical counterpart,
 are found to be at the
core of intermediate statistics. The system is interacting, but
no classical forces act on the particles, the charge $e$ and the
flux $\phi$ are
statistical. Note, however, that in dimension 1,
the Calogero model [4], with $1/x_{ij}^2$ interactions,
is also considered as an intermediate statistics model.
Classical forces do act on the
particles, but one finds that their asymptotic properties are not
affected by the interaction, up to a permutation.

The $A-B$ effect being historically one of the first examples of a mesoscopic
effect, it is quite appropriate to discuss its $N$ body generalization,
i.e. the anyon model, in the present volume.

As already emphasized above, a direct consequence of the singular
gauge transformation (2) is that
particles carry flux tube where an infinite statistical magnetic field
(a pseudo-scalar in 2+1 dimensions) is
concentrated
\be\label{B} B=\phi\sum_i\delta(\vec r-\vec r_i)=\phi\rho\ee
where $\rho$ is the density of particles.
This is one hallmark of the model : the magnetic field is entirely
defined by the distribution of particles, there is no intrinsic degree
of freedom -a propagating photon- associated to the gauge field.
Again, this is nothing but saying that for a statistical (topological)
interaction, no real
electromagnetic forces should enter the game. Therefore, in a Lagrangian
formulation,
one does not expect a Maxwell
term, but rather a topological term, to define the intrinsic dynamic of
the statistical gauge field. In 2+1 dimension, the Lagrangian
\be L_N=\sum_{i=1}^N {1\over 2}mv_i^2 + e(\vec A_i.\vec v_i-A_0(\vec
r_i))+{\kappa\over 2}\epsilon_{\mu\nu\rho}\int d^2\vec r
A^{\mu}\partial^{\nu}A^{\rho}
\ee
minimally couples $N$ free classical particles to a Chern-Simons gauge
field $A^{\mu}$. The Chern-Simons term,
$\epsilon_{\mu\nu\rho}A^{\mu}\partial^{\nu}A^{\rho}$,
is gauge invariant (up to a total divergence),
is topological (there is no metric), and breaks parity.
Writing the equation of motion
w.r.t. the time component $A^0$ field, one finds
$e\rho=\kappa B$,
that is to say (\ref{B}) with $e/\kappa=\phi$. It is not difficult to
solve this equation for $\vec {A}(\vec r)$ in the Coulomb gauge, and,
first-quantizing
the model, to get the desired Hamiltonian (\ref{H}).
Note that in 3+1 dimension the topological term
$\epsilon_{\mu\nu\rho\lambda}F^{\mu\nu}F^{\rho\lambda}$ is a
total divergence -actually the divergence of a Chern-Simons term- and
does not contribute to the equations of motion. One concludes again that
it is only in 1+1 or 2+1 dimensions that non trivial statistics can be
properly defined.

What can be done exactely in the anyon model ? Not so much, in fact,
even if important efforts have been devoted to the study of the N-anyon
spectrum.
Let us focus on the solvable 2-anyon problem -solvable
because the relative
problem
is nothing but the original $A-B$ problem. In the notations
$r_{12}=r$, and $\theta_{12}=\theta$,
the relative Hamiltonian
\be \label{8} mH_{rel}   =  -\partial  _r^2  - {1\over  r}\partial  _r
-{1\over
r^2}\partial _{\theta}^2 + 2i{\alpha  \over r^2}\partial _{\theta}
+{\alpha ^2\over r^2} \ee
when acting on the relative
wavefunction   $\psi(r,\theta)=\exp(im\theta)f(r)$
-$m$ even-  yields the radial Schr\"odinger
equation
$mH_{rel}=-\partial  _r^2 - {1\over r}\partial _r +
{(m-\alpha)}^2{1\over r^2} $,
where the shifted angular momentum $(m-\alpha)$ appears explicitely.
The spectrum $E=k^2/m$ of $H_{rel}$ is continuous
with normalized states in the continuum
$\psi(r,\theta)=\exp(im\theta){ \sqrt {k/2\pi}} J_{|m-\alpha|}(kr)$.
Note that $\exp(im\theta){\sqrt{k/2\pi}}J_{-|m-\alpha|}(kr)$ is also solution,
but not normalizable, except
if $m=0$. In principle, one should consider s-wave linear combinations
of both solutions, defining a one parameter family of self-adjoint
extensions. They are, however,
 forbidden because
$J_{-|\alpha|}$ diverges at coinciding points
 $r\to 0$. But, in quantum mechanics, exclusion of the diagonal of
the  configuration space
means that only
solutions  satisfying a
particular choice of boundary conditions at the origin
-they vanish,  one speaks of hard-core anyons- should be retained.

The 2-anyon relative partition function
diverges  as the volume. In order to give a unambiguous meaning to the
volume for
thermodynamical quantities, such as the equation of state, one
has to regularize the system at long distance, and then take
the thermodynamic limit. In the anyon model, one confines
the particles in a harmonic potential, by adding
$\sum_i{1\over 2}m\omega^2r_i^2$ to the Hamiltonian,
to get, up to a normalisation
\be \label{10}\psi^{\alpha}_{nm}= e^{im\theta}e^{-m\omega r^2/4}
r^{\vert  m-\alpha
\vert/2}L_n^{\vert  m-\alpha
\vert}(m\omega r^2/2)\ee
  The discrete spectrum
$E_{nm}= (2n+\vert  m-\alpha \vert +1)\omega$
 interpolates linearly between the Bose and Fermi spectra when $\alpha
\quad :
0\to \pm 1$.
One notes that, when $m=0$, the spectrum is nonanalytic in $\alpha$.

What about the $N$-anyon problem ? There exits  $N$-anyon linear states
that generalize the
2-anyon
linear states (\ref{10}),  but they are
only a part of the spectrum. This can already be seen for
the 3-anyon groundstate.
Numerical as well as
perturbative numerical studies indicate
that a nonlinearly interpolating  state
departs from the  3-fermion
groundstate, but its exact form is still unknown.
To proceed further, three possible routes can be followed :

\noindent i) make some
approximation, as the mean magnetic field approximation

\noindent ii) perturb [5] around Bose and
Fermi statistics with a small anyonic interaction

\noindent iii) simplify [6]
drastically
the
system, by projecting it on the lowest landau level (LLL)
 of an external
strong magnetic field, at low temperature. Far from being academic,
this projection  corresponds to the fractional quantum Hall regime.
On the theoretical side, it is interesting since  it amounts to consider a one
dimensional version of the anyon model, which is intimately related to the
Calogero
model.

\noindent i) What is exactely meant by a mean field approximation ? Just look
at
(\ref{B}) and
consider that the $N^{th}$ anyon is subject to the
average magnetic field $<B>$ due to the $N-1$ other anyons
\be\label{MF} <B>V\simeq \phi(N-1) \ee
which should be understood in the thermodynamic
limit $N\to \infty $,  $V\to \infty$,  $\rho=N/V$.
 In the case where an external
magnetic field $B$ is present,
this approximation is valid if the classical cyclotron orbit of a
given anyon  encloses a big number of flux tubes, that is to say if
$\rho>>\rho_L+\rho\alpha$, where $\rho_L=eB/2\pi$ is the Landau
degeneracy  of the external $B$ field (without loss of generality one
has assumed $eB>0$). It means  that the appropriate average field
limit is $\rho\to\infty,\alpha\to 0$, with $\rho\alpha\phi_o=<B>$ finite.
In these conditions, each anyon
feels a
total magnetic field $B+<B>$, and has
$d=N_L+\alpha (N-1)$ one-body quantum states available in a given Landau
level of the
total magnetic field. Suppose now one restricts to the LLL
-again the Hall regime : one has $N$ bosons,
to distribute in $d$ quantum
states, the number of possibilities being
\be\label{d} {(d+N-1) !\over N ! (d-1) !}\ee
If $\alpha\in[-1,0]$ -the flux tubes carried by the anyons are
antiparallel to the external magnetic field- one finds that
(\ref{d}) interpolates between the Bose
and Fermi ends. One has here a good example of
Haldane's exclusion statistics concept [7], which is
defined via Hilbert space counting argument : when one adds extra
anyon to the system, from Haldane's definition  one has that ${\Delta d/\Delta
N}=-g$,
where $g=-\alpha$ is the exclusion statistical parameter. $g=0$ for
bosons, since adding one extra boson does not change the number of
quantum states available for the other bosons (Bose condensation),
whereas $g=1$ for fermions, since adding one
fermion diminishes by 1 the number of quantum states available for the
other fermions (Pauli exclusion). We will come back later on the
thermodynamical properties of the anyon gas in the LLL of an external
$B$ field.

ii) The fact that a perturbative approach is possible is not
obvious. The non analytical $|\alpha|$ behavior of the $m=0$ eigenstates in
(\ref{10})
is a clear indication of the failure of standard
perturbation.
Put differently, the unperturbed
Hilbert space $\psi_{nm}^0
(r, \theta )
\simeq r^{\vert m\vert}, r\to 0$, is not adapted
to the domain of
definition of the Hamiltonian,  since the s-wave
states do
not vanish at
the
origin
contrary to the exact s-wave states which do vanish as
$\psi_{n0}^{\alpha}
(r, \theta )\simeq
r^{|\alpha|}$.
If one estimates the matrix element
 \be \label{13}\ba {ll} \alpha <\psi_{nm}^0
(r, \theta)\vert {2i\over r^2}\partial_{\theta} \vert
\psi_{nm}^{0} (r, \theta )>&=-{m\over \vert m\vert}\alpha\omega \  \ {\rm for}\
m\ne 0 , \\
 &= 0 \ \ \ {\rm for}\  m=0\ea \ee
one finds obviously no contribution
 from the s-wave states. However, perturbation theory makes sense
only when all perturbative corrections to the zeroth order spectrum are
finite,
in particular $\alpha^2<\psi_{nm}^0
(r, \theta)\vert 1/r^2 \vert
\psi_{nm}^0 (r, \theta)>
\to \alpha^2 \int dr  r^{2\vert m\vert -1}$.
For $m \ne 0$, this is the case, but for $m=0$, it
is logarithmically divergent.
As already
noticed when disregarding possible self-adjoint extansions,
the   s-wave states short distance behavior
has to be treated with a  particular care. If one redefines
\be \label{redef}\psi(r,\theta)= r^{|\alpha|}\tilde\psi(r,\theta),\ee
one finds that not only
the singular
${\alpha^2/ r^2}$ term disappears from the Hamiltonian $ m{\tilde {H}}_{rel}
=  -\partial  _r^2  - {1\over  r}\partial  _r
-{1\over
r^2}\partial _{\theta}^2 + 2i{\alpha  \over r^2}\partial _{\theta}
-2\vert \alpha \vert {\partial_r\over r}$ acting on
$\tilde \psi(r,\theta)$,
but also that the new $\vert \alpha \vert$ term is perfectly suited
for a perturbative analysis.
Indeed, the exact 2-anyon spectrum (\ref{10})
is recovered from $\tilde{H}$  at first order in $\alpha$ and $|\alpha|$,
all
higher order terms being finite and  cancelling !

In the N-anyon case, one simply generalizes
by $\psi(\vec r_1,\cdots,\vec r_N)= \prod_{i<j} r_{ij}^{{|\alpha|
}}
 \tilde{\psi}(\vec r_1,\cdots,\vec r_N)$.
Not only the $\alpha^2/r_{ij}^2$ 2-body singular terms
disappear in the new Hamiltonian acting
on $\tilde{\psi}$,
\be\label{17}  \tilde {H}_N=\sum _{i=1}^{N}({\vec{p}_i ^2\over
2m}+{i\alpha\over m}\sum_{j\ne i}{\vec k \times \vec r_{ij}\over r^2_{ij}}\vec
\partial_i
-{\vert \alpha\vert\over m}\sum_{j\ne i}{\vec r_{ij}
\over r^2_{ij}}\vec \partial_i),\ee
but also the 3-body terms.
This singular perturbative algorithm, and the resulting Hamiltonian
(\ref{17}),
have been used for
computing the equation of state of an anyon gaz at second order in
perturbation theory. In fact, it amounts to introduce an
additional spin-like degree of freedom for each anyon. To see this,
one notes  that
\be\label{18} {H'_{N}}={H_{N}}
+\sum_{i<j}{2\pi|\alpha|\over m_o}\delta(\vec r_{ij})\ee
is in fact the  Hamiltonian from which $\tilde{H}_N$ is derived via the
wavefunction redefinition (\ref{redef}), when one properly pays attention to
possible Dirac distributions coming from the action of the Laplacian on
$|\alpha|\ln r_{ij}$. The Hamiltonian $H'_N$
leads to the same perturbative expansion as $\tilde{H}_N$, in particular
the perturbative divergences due to $\alpha^2/r^2_{ij}$ exactly cancel
those due to the contact term. These $\delta^2(\vec r_{ij})$ short range
repulsive
interactions are needed to implement the exclusion of the diagonal
of the configuration space (hard-core anyons). On the other hand,
the Hamiltonian (\ref{18}) is
a
particular Aharonov-Casher Hamiltonian, in which each anyon
is coupled to the infinite magnetic
field insides the flux tubes carried by the other anyons,
by a magnetic moment $\mu=-{e\over 2m}{|\alpha|\over \alpha}$.
In that sense, hard-core anyons have to be considered as
spin 1/2 like non relativistic objects.

iii) Let us finally couple the model to a
strong external magnetic field $\vec B
=B\vec k$
\be \label{BB}  H_N=\sum_{i=1}^N  {1\over 2m}
               \bigg(\vec{p}_i -\alpha\sum_{j\ne i}
               {\vec k\times\vec r_{ij}\over r_{ij}^2}-e{B\over 2}\vec k
               \times\vec r_i\bigg)^2   \ee
The sign of $\alpha$ is now a physical
observable, implying that the interpolation
$\alpha\in[-1,1]$
is not symmetric with respect to $\alpha=0$, i.e. things are
not the same when $\alpha\to 0^+$  or $\alpha\to 0^-$.
One can illustrate this statement on the groundstate with
energy $N\omega_c$
\be\label{BBB}  \psi=z^{l} \prod_{i<j} r_{ij}^{-\alpha}
                   \prod_{i<j} z_{ij}^{m_{ij}}
                   \exp(-{m\omega_c\over 2} \sum_i^Nz_i\bar z_i)
                   \quad\quad l\ge 0 \quad m_{ij}\ge\alpha
\ee
where $z=\sum_i z_i/N$ is the center of mass coordinate
($z_i=x_i+iy_i$).
If $\alpha\in [-1,0]$, then $m_{ij}\ge 0$, the groundstate basis
(\ref{BBB})
is complete, and yields  when $\alpha \to 0^-$ the complete
$N$-bosons LLL basis. On the other hand, when $\alpha \in [0,1]$, then
$m_{ij}>0$, the groundstate basis is incomplete, and the limit
$\alpha\to 0^{+}$ yields only part of the   Landau basis :
some excited states (in fact an infinity), which are not
analytically known, merge in
the grounsdate when $\alpha\to 0^{+}$.
In order to have a cyclotron gap of order $2\omega_c$ above
the $N$-anyon groundstate (\ref{BBB}),
such that the thermal probability to have
an excited state $e^{-2\beta\omega_c}$ is negligible
 when the thermal energy $1/\beta$
is smaller than the gap,
one has to constrain $\alpha\in [-1,0]$,
since it is only in this interval that a continuous
interpolation between the
bosonic and fermionic groundstates does not require
unknown non linear excited states.
One can easily convince oneself of what has just been said simply by
extracting from (\ref{BBB})
the anyonic prefactor
$\prod_{i<j} r_{ij}^{|\alpha|} $,
to get, when $\alpha\in [-1,0]$,  the product
$\{ {\displaystyle \prod_{i=1}^N} z_i^{\ell_i}
\exp(-{1\over 2}m\omega_c z_i{\bar {z}}_i), \ell_i\ge 0\}$
of 1-body Landau
groundstates
of energy $\omega_c$ and angular momentum $\ell_i$.
Extracting this prefactor amounts to go in the $\tilde{\quad}$
representation. When
projected on the LLL basis,  $\tilde{H}_N$ becomes a sum of 1-body Landau
Hamiltonian
\be \label{cal}\tilde{H}_N= \sum_iH_B(\vec r_i)-\alpha
{N(N-1)\over 2}(\omega_t-\omega_c)\ee
shifted by the constant
 $N\omega_c\to N\omega_c-\alpha N(N-1)(\omega_t-\omega_c)/2$,
which partially lifts the groundstate degeneracy. A
harmonic regulator $\omega$ has been introduced, such that
$\omega_t=\sqrt{\omega_c^2+\omega^2}\to \omega_c $ when $\omega\to 0$.
(\ref{cal}) is indeed diagonal $N\omega_c$ in
the thermodynamic limit $\omega\to 0$.
 On the other hand, when $\alpha\in [0,1]$,
identical manipulations lead to
\be \tilde{H}_N= \sum_iH_B(\vec r_i)+\sum_{i<j}{4\pi\alpha\over
m}\delta(\vec r_{ij})-\alpha {N(N-1)\over 2}(\omega_t-\omega_c),\ee
that is to say, in the thermodynamic limit,
 to a sum of $\delta$ interactions which
partially encode the excited states joining the groundstate when
$\alpha\to 0^{+}$.
Let us concentrate on the thermodynamic of  (\ref{cal}).
The spectrum is quite reminiscent of whose of the Calogero
model in a harmonic well. In the thermodynamic limit,
the thermodynamical potential is found to be
\be\label{16}  \Omega\equiv -\sum_{N=1}^\infty b_N z^N
                    = -V\rho_L\ln y(ze^{-\beta\omega_c})  \ee
where $y(z')$ is solution of $y-z' y^{\alpha+1}=1$ with $y(z')\to 1$
when $z'\to 0$.
The filling factor, $\nu\equiv \rho/\rho_L$, is found to be
implicitely defined by
$y(ze^{-\beta\omega_c})=1+\nu/(1+\alpha\nu)$. $\nu$
gives the number of
anyons one can put in a given LLL quantum state, and thus
is nothing but an occupation number, which is found to be constrained by
\be
ze^{-\beta\omega_c}={\nu\over(1+\nu(1+\alpha))^{1+\alpha}
(1+\alpha\nu)^{-\alpha}},\ee
This equation is central to the thermodynamic of the
Calogero model and the anyon model in a strong B field, and to
Haldane's exclusion statistics  as well.
The equation of state reads
\be\label{P}  P\beta=\rho_L
                      \ln\bigg(1+{\nu\over{1+\alpha\nu}}\bigg) \ee
correctly interpolates, when $\alpha \in [-1,0]$, between the Bose
and Fermi equations of state in the LLL. In the Bose case, $\alpha=0$,
any $\nu$
is allowed
because of Bose condensation.
In the Fermi case, $\alpha=-1$,
Pauli exclusion implies that the lowest Landau level
is completly filled when $\nu=1$.
At an intermediate $\alpha$, the critical filling $\nu_{cr}=-1/\alpha$
can be interpreted as at most $-1/\alpha$ anyons of statistics $\alpha$
can occupy a given LLL quantum state.
Since transitions to excited levels are by construction
forbidden, the pressure diverges when the
lowest Landau level is fully occupied such
that any additional particle is excluded.
In this situation the gas is incompressible, and the groundstate is
nondegenerate with the $\ell_i$'s all nul
$\psi_{\alpha}=\prod_{i<j} r_{ij}^{-\alpha}
                   \exp(-{m\omega_c\over 2} \sum_i^Nz_i\bar z_i)$.
In the fermionic case $\alpha=-1$,
one indeed recovers, in the singular gauge, a Vandermonde
determinant.
At the critical filling,
the magnetic field is entirely screened by the anyon gas (remember that
each anyon carries
$\alpha \phi_o$ flux). This is quite analogous to what happens in the
FQHE, where the Laughlin wavefunctions [8] at filling $\nu=1/m$ are such as
each electron carries $m\phi_o$ flux, so that the external $B$ field is
screenned by the electron gas. As a final side-remark, if one considers
$\psi_{\alpha}$ in the
singular gauge, and  analytically continuates $-\alpha\to m$, one recovers
the non degenerate Laughlin wavefunctions for the incompressible quantum
Hall fluid, at fractional filling $1/m$.

{\bf References }

1. J. M. Leinaas and J. Myrheim,  Nuovo  Cimento B {\bf 37} (1977) 1

2. for a review on the Anyon model, see
 A. Lerda, "Anyons", Springer- Verlag (1992)

3. Y. Aharonov and D. Bohm, Phys. Rev. {\bf 115} (1959) 485

4. F. Calogero, J. Math. Phys. 10 (1969) 2191 and 12 (1971) 419

5. for a review on the perturbative approach in the anyon model, see
S. Ouvry, Phys. Rev. D 50 (1994) 5296, and references therein ; see also A.
Comtet, S. Mashkevich and S.
Ouvry, "Magnetic Moment and Perturbation Theory in 2-d Gauged Quantum
Mechanics" IPNO/TH 94-91

6. for a review on the thermodynamic of an anyon gas in a strong $B$
field and its relation with the Calogero model, see A. Dasni\`eres de Veigy
and S. Ouvry, "One dimensional Statistical Mechanics for Identical
Particles : the Calogero and Anyon Cases", cond-mat 9411036, International
Journal of Mod. Phys.
Lett. B (to be published), and references therein

7. F. D. M. Haldane, Phys. Rev. Lett. 67 (1991) 937 ; Y. S. Wu, Phys.
Rev. Lett. 73 (1994) 922

8. R. B. Laughlin,    Phys. Rev. Lett. {\bf 50} (1983) 1395

\end{document}